\documentclass[onecolumn,draft]{IEEEtran}

\usepackage{amsfonts}
\usepackage{makeidx}
\usepackage{amssymb}
\usepackage{graphicx}
\usepackage{amsmath}
\usepackage{fancyhdr}
\usepackage{curves}
\usepackage{pdfpages}
\usepackage{latexsym}
\usepackage{amsthm}
\usepackage{tikz-cd}

\newtheorem{theorem}{Theorem}
\newtheorem{definition}[theorem]{Definition}
\newtheorem{lemma}[theorem]{Lemma}
\newtheorem{proposition}[theorem]{Proposition}
\newtheorem{corollary}[theorem]{Corollary}

\newtheorem{remark}[theorem]{Remark}
\newtheorem{ex}[theorem]{Example}

\begin{document}

\title{Vard{\o}hus Codes: Polar Codes Based on Castle Curves Kernels}

\author{
	\IEEEauthorblockN{Eduardo Camps\IEEEauthorrefmark{1}\thanks{\IEEEauthorrefmark{1} Departamento de Matem\'aticas -- Instituto Polit\'ecnico Nacional -- M\'exico --
			Email: ecfmd@hotmail.com 
			-- Partially supported by a grant 'Beca Mixta', CONACYT (M\'exico)}},
	\and
\IEEEauthorblockN{Edgar Mart\'inez-Moro\IEEEauthorrefmark{2}\thanks{\IEEEauthorrefmark{2} Institute of Mathematics -- University of Valladolid -- Castilla, Spain -- Email: Edgar.Martinez@uva.es -- Partially funded by Spanish-MINECO MTM2015-65764-C3-1-P research grant.}},
\and and
\IEEEauthorblockN{Eliseo Sarmiento\IEEEauthorrefmark{3}\thanks{\IEEEauthorrefmark{3} Departamento de Matem\'aticas -- Instituto Polit\'ecnico Nacional -- M\'exico -- Email: esarmiento@ipn.mx -- Partially supported by SNI--SEP.}}
}

\markboth{Castle curves and polar codes}%
{Shell \MakeLowercase{\textit{et al.}}: Bare Demo of IEEEtran.cls for IEEE Journals}

\maketitle

\begin{abstract}
In this paper we show some applications of algebraic curves to the construction of kernels of polar codes over  a
discrete memoryless channel which is symmetric w.r.t the field operations. We will also study  the minimum distance of the  polar codes proposed, their duals and the exponents of the matrices used for defining them. All the restrictions that we make to our curves will
be accomplished by the so called Castle Curves.
\end{abstract}


\IEEEpeerreviewmaketitle

\section{Introduction}

\IEEEPARstart{G}{iven} a non-singular square matrix $G$ of size $l\times l$ over the finite field  $\mathbb{F}_q$  with $q$ elements ($q$ a power of a prime) it can be used for designing the kernel of a polar code. In such construction it is interesting to study the exponent of the matrix and the information set.

Concerning the information set, it was proved in \cite{Bardet} that given a binary symmetric channel the construction based on the binary matrix $G_A=\begin{bmatrix} 1&0\\1&1\end{bmatrix}$ can be analyzed in terms of the elements in the polynomial ring  $\mathbb{F}_2[x_1,\ldots,x_n]/\langle x_1^2-x_1,\ldots,x_n^2-x_n\rangle$. That is,  for a fixed given monomial order (thus giving a divisibility on the ring and a weight to the variables  $x_i, \, i=1,\ldots, n$) one can get information for its information set. In the same paper the authors also devise a formula for the minimum distance, derived the dual code and they proved that the permutation group of the code was ``large''.

In this paper we will show  some applications of algebraic curves to polar codes. If we apply some restrictions to a discrete memoryless channel (DMC)  
 $W:\mathbb{F}_q\rightarrow\mathcal{O}$ and we will study under which assumptions a matrix  $G$ polarizes in terms of the curve used to construct its kernel.   Also the information set, the minimum distance of a polar code and its dual based on the curve will be shown generalizing some of the results in \cite{Bardet}, in particular we will keep the same notation they used for some properties of polar codes. All the restrictions that we  make to our curves will be accomplished by Castle Curves \cite{castillo1}, this is the reason for the title since Vard{\o}hus (Norway) is the only castle known by the authors in the polar region.
 
 The structure of the paper is as follows.  In Section~\ref{sec:pre} we have compile some basic facts on polar codes and algebraic geometric curves needed to understand the paper.  Section~\ref{sec:AGkernels} reviews some results in   \cite{Anderson}  and we adapt them for constructing kernel matrices that arise from algebraic curves for a SOF channel that is a discrete memoryless channel which is symmetric w.r.t the field operations. Section~IV deals with the computation of the minimum distance and the dual of codes proposed in the previous section. Finally Section~V is devoted to the study of the exponent of such codes and how to reduce them to get a better  exponent for a given matrix size.

\section{Preliminaries}\label{sec:pre}

\subsection{Polar codes}
Ar\i kan introduced in \cite{Arikan} a method to get efficient capacity-achieving binary source and channel codes, generalized later by \c Sa\c so\u glu et. al. in \cite{Sasoglu}. In this section, we review briefly some results used in the rest of the work. 
Given a non-singular matrix $G$ over $\mathbb{F}_q$ ($q=p^r$ a power of a prime) of size $l\times l$ and a discrete memoryless channel $W:\mathbb{F}_q\rightarrow\mathcal{O}$, take $N=l^n$ and define $W_n:\mathbb{F}_q^N\rightarrow\mathcal{O}^N$ such that
$$W_n(y_1^N\ |\ u_1^N)=\prod_{i=1}^N W(y_i\ |\ u_1^N G_n)$$
with $G_n=B_n G^{\otimes n}$, where $G^{\otimes n}$ is the Kronecker product of $G$ with itself $n$ times and $B_n$ is the matrix $N\times N$ such for the vector $v_1^N=u_1^NB_n$ yields that if $(i_n,\ldots,i_1)$ is the $l$-ary expansion of $i$, then $v_i=u_{i'}$, where $i'$ has the expansion $(i_1,\ldots,i_n)$. The channel $W_n$ is splitted into $N$ channels $W_n^{(i)}:\mathbb{F}_q\rightarrow\mathcal{O}^{N}\times\mathbb{F}_q^{i-1}$ with
$$W_n^{(i)}(y_1^N,u_1^{i-1}\ |\ u_i)=\sum_{u_{i+1}^N\in\mathbb{F}_q^{N-i}}W_n(y_1^N\ |\ u_1^N)$$

These channels are compared via its rate, that is, for a channel $W:\mathbb{F}_q\rightarrow\mathcal{O}$
$$I(W)=\sum_{y\in\mathcal{O}}\sum_{x\in\mathbb{F}_q}\frac{1}{q}W(y|x)\log_q\left(\frac{W(y|x)}{W_{\mathcal{O}}(y)}\right)$$

\noindent where $W_{\mathcal{O}}(y)$ represents the probability of receive $y$ through $W$. We say that $G$ polarizes $W$ if for each $\delta\in(0,1)$ we have:
\begin{align*}
&\lim_{n\rightarrow\infty}\frac{\left|\left\{i\in\{1,\ldots,N\}\ |\ I\left(W_n^{(i)}\right)\in(1-\delta,1]\right\}\right|}{N}=I(W)\\
\\
&\lim_{n\rightarrow\infty}\frac{\left|\left\{i\in\{1,\ldots,N\}\ |\ I\left(W_n^{(i)}\right)\in[0,\delta)\right\}\right|}{N}=1-I(W)
\end{align*}
If a matrix $G$ polarizes, we say that $G$ the kernel of the polarization. Ar\i kan's original construction using $G_A=\begin{bmatrix} 1&0\\ 1&1\end{bmatrix}$ polarizes any binary symmetric channel $W$. Defining a channel over $\mathbb{F}_q$ as symmetric if for each $x\in\mathbb{F}_q$ exists a permutation $\sigma_x$ such $W(y|x)=W(\sigma_{x'-x}(y)|x')$ for each $y\in\mathcal{O}$ and $x',x\in\mathbb{F}_q$, Mori and Tanaka~\cite{moriq} showed that source polarization is the same as symmetric channel polarization and

\begin{theorem}[\cite{moriq}, Theorem 14]	
	Let $G$ be a non-singular matrix of size $l\times l$ over $\mathbb{F}_q$, $V$ an invertible upper triangular matrix and $P$ a permutation matrix. If $G'=VGP$ is a lower triangular matrix with units on its diagonal (we call it a standard form of $G$), then for a $G$ with a non-identity standard form the following statements are equivalent	
	\begin{itemize}
		\item Any symmetric channel is polarized by $G$.
		
		\item $\mathbb{F}_p(G')=\mathbb{F}_q$ for any standard form $G'$ of $G$, where $\mathbb{F}_p(G')$ denotes the field generated by the adjunction of the elements in $G'$.
		
		\item $\mathbb{F}_p(G')=\mathbb{F}_q$ for one standard form $G'$ of $G$.
	\end{itemize}
\end{theorem}

When a matrix $G$ polarize some channel $W$ we get a efficient codes choosing the best channels $W_n^{(i)}$. For this purpose, we use the Bhattacharyya parameter  for a channel $W:\mathbb{F}_q\rightarrow\mathcal{O}$ defined as follows
$$Z(W)=\frac{1}{q(q-1)}\sum_{x,x'\in\mathbb{F}_q, x\neq x'}\sum_{y\in\mathcal{O}}\sqrt{W(y|x)W(y|x')}.$$

We construct a polar code choosing an information set $\mathcal{A}_n\subset\{1,\ldots,N\}$ with the condition that for each $i\in\mathcal{A}_n$ and $j\notin\mathcal{A}_n$ yields
$$Z\left(W_n^{(i)}\right)\leq Z\left(W_n^{(j)}\right).$$

The polar code $C_{\mathcal{A}_n}$ is generated by the rows of $G_n$ indexed by $\mathcal{A}_n$. Due to the polarization of $G$ over $W$, this code will have a low block error probability. To see this, we use the concept of rate of polarization or exponent of $G$ introduced by Korada et al. in \cite{korada} and generalized by Mori and Tanaka in \cite{morinl}. The exponent of $G$ is defined as
$$E(G)=\frac{1}{l\ln l}\sum_{i=1}^l\ln D_i,$$
where $D_i$ is called the partial distance and it is define as $D_i=d(G_i,\langle G_{i+1},\ldots,G_l\rangle)$, where $G_i$ is the $i$-th row of $G$.  The original Ar\i kan's matrix $G_A$ has exponent $\frac{1}{2}$. The exponent is the value such that
\begin{itemize}
	\item for any fixed $\beta<E(G)$
	$$\liminf_{n\rightarrow\infty} P[Z_n\leq 2^{-N^\beta}]=I(W).$$
	
	\item For any fixed $\beta>E(G)$
	$$\liminf_{n\rightarrow\infty} P[Z_n\geq 2^{-N^\beta}]=1,$$
\end{itemize}
where $Z_n=Z(W'_n)$ and $W'_n={W'_{n-1}}_1^{(B_n)}$ with $\{B_n\}_{n\in\mathbb{N}}$ independent random variables identically distributed over $\{1,\ldots,l\}$.

Anderson and Matthews proved in \cite{Anderson} that this means that for any $\beta<E(G)$  polar coding using kernel $G$ over a   DMC channel $W$ at a fixed rate $0<R<I(W)$ and block length $N=l^n$ implies
$$P_e=O(2^{-N^\beta}),$$
where $P_e$ is the probability of block error.

The partial distances $D_i$ can be estimated by the sucesion of nested codes $\langle G_i,\ldots, G_l\rangle$ and shortening matrices leads to a good exponents in smaller sizes. Compute the exponent and the information set $\mathcal{A}_n$ are two of the main problems in polar coding. About the last topic Bardet et al.~\cite{Bardet} proved that for $G_A$ the structure of the information set can be derived from a monomial order over $\mathbb{F}_2[x_1,\ldots,x_n]$ and they proved also that minimum distances are computable and duals of polar codes have similar structures using the fact that rational curves over $\mathbb{F}_2$ have nice properties.
All these conditions leads to consider a special type of algebraic curves, the Castle-like curves that have a nested code structure. They can be described in terms of a finite-generated algebra and satisfied the isometry-dual property.

\subsection{Algebraic pointed curves and AG codes}

Let us remember some facts about algebraic geometry (AG) codes over curves (for an extensive account on AG codes see for example~\cite{AGC}). By a curve we mean a projective, non-singular, geometrically irreducible algebraic curve $\mathcal{X}$ over $\mathbb{F}_q$ and we denote by $\mathcal{X}(\mathbb{F}_q)$ its rational points, by $\mathbb{F}_q(\mathcal{X})$ its function field and by $g=g(\mathcal{X})$ its genus. We will consider two rational divisors $D=\sum_{i=1}^l P_i$, where the $P_i$, $i=1,\ldots , l$ are distinct rational points in the curve (rational places) and $G$ such that $\mathrm{supp}\ D\cap\mathrm{supp}\ G=\emptyset$ and $1\leq \mathrm{deg}(G)\leq n+2g-1$. We define the evaluation map $ev_D:\mathcal{L}(G)\rightarrow\mathbb{F}_q^l$ as
$$ev_D(f)=(f(P_1),\ldots,f(P_l)),$$
where $\mathcal{L}(G)$ is the vector space of rational functions over the curve such that either $f=0$ or $\mathrm{div}(f)+G\geq 0$.
 We define the evaluation code as $C(D,G)=ev_D(\mathcal{L}(G))$. The kernel of $ev_D$ is $\mathcal{L}(G-D)$ and   the length of $C(D,G)$ is $\deg D$, its dimension $k=l(G)-l(G-D)$ and its minimum distance $\delta(C(D,G))\geq \deg D-\deg G$.

 Given $Q\in\mathcal{X}(\mathbb{F}_q)$ we called a pointed curve to the pair $(\mathcal{X},Q)$. We denote by $H(Q)$ to the Weierstrass semigroup of $Q$ and given $D$ as before we denote $$H^\ast(Q)=\{m\in\mathbb{N}_0\ |\ C(D,(m-1)Q)\neq C(D,mQ)\}.$$ Clearly, $|H^\ast(Q)|=l$ and we can write $H^\ast(Q)=\{m_1,\ldots,m_l\}$. Most of information of the codes $\{C(D,mQ)\}_{m\in\mathbb{N}_0}$ is contained in $H^\ast(Q)$.

If $\mathcal{X}$ is a curve of genus $g$ we say that $H(Q)$ is symmetric if
$$h\in H(Q)\Longleftrightarrow 2g-1-h\notin H(Q).$$
When $H(Q)$ is symmetric and $D\equiv lQ$ we have $H^\ast(Q)=H(Q)\cap\{0,1,\ldots,n-1\}\cup\{l_1,\ldots,l_g\}$ where $l_i$ are gaps of $Q$~\cite{diegorder}. The isometry-dual condition for a sequence of codes of length $l$ $\{C_i\}_{i=1}^l$, $C_i\subsetneq C_{i+1}$, means that there is $x\in\mathbb{F}_q^l$ such that for each $i\in\{1,\ldots, l\}$, $C_i^\perp$ is isometric by $x$ to $C_{l-i}$. In \cite{diegorder} they also proved that the following statements are equivalent when $l \geq 2g-2$
\begin{itemize}
	
	\item The set $\{C(D,mQ)\}_{m\in H^\ast(Q)}$ satisfies the isometry-dual condition,
	
	\item the divisor $(l+2g-2)Q-D$ is canonical,
	
	\item $l+2g-1\in H^\ast(Q)$.
	
\end{itemize}
Then if $l\geq 2g-2$ and $D\equiv lQ$, the sequence of nested codes $\{C(D,m_iQ)\}_{m_i\in H^\ast(Q)}$ satisfy the isometry dual condition. Observe that a rational curve satisfies these conditions.


\section{Algebraic Curves and Kernels}\label{sec:AGkernels}
From now on $G$ will be  a non-singular square matrix $G$ of size $l\times l$ over the finite field  $\mathbb{F}_q$  with $q$ elements ($q=p^r$ a power of a prime)  and  $W:\mathbb{F}_q\rightarrow\mathcal{O}$ a  DMC channel. Let $G_n$ be the matrix used for constructing a polar code of length $N=l^n$ based on $G$ and consider $$W_n(y_1^N|u_1^N)=\prod_{k=1}^N W(y_k|u_1^l(G_n)_{\ast,k})$$
and the partitions given by
$$W_n^{(i)}=(y_1^N,u_1^{i-1}|u_i)=\sum_{u_{i+1}^N} W_n(y_1^N|u_1^N).$$
Note that the channels 
  $\left(W_{n-1}^{(i)}\right)^{(j)}$ and $W_n^{(i-1)l+j}$ are the same in the sense that the parameters $I(W)$ y $Z(W)$ are equal in both cases.

\begin{proposition}\label{degchanpart}
If  $W:\mathbb{F}_q\rightarrow\mathcal{O}$ is a  DMC given two integers  $1\leq i\leq l^{n-1}$ and $1\leq j\leq l$ then
	$$\left(W_{n-1}^{(i)}\right)_1^{(j)}= W_n^{((i-1)l+j)}.$$
\end{proposition}

\begin{IEEEproof}
	Let  $$f:\mathcal{O}^{l^n}\times\mathbb{F}_q^{l(i-1)}\rightarrow\left(\mathcal{O}^{l^{n-1}}\times\mathbb{F}_q^{i-1}\right)^l$$ defined as follows
	$$f(y_1^{l^n},u_1^{l(i-1)})=(y_{(k-1)l^{n-1}+1}^{kl^{n-1}},u_1^lG_{\ast,k},\ldots,u_{(i-2)l+1}^{(i-1)l}G_{\ast,k})_{k=1}^l.$$
Then we have that 
	\begin{align*}
	&W_n^{((i-1)l+j)}(y_1^{l^n},u_1^{(i-1)l+j-1}|u_j)\\=&\frac{1}{q^{l^{n}-1}}\sum_{u_{(i-1)l+j+1}^{l^n}}\prod_{k=1}^{l^n}W\left(y_k\left|\sum_{h=1}^{l^n} u(G_n)_{\ast,k}\right.\right)\\
	=&\frac{1}{q^{l^{n}-1}}\sum_{u_{(i-1)l+j+1}^{l^n}}\prod_{k=1}^{l}\prod_{k'=1}^{l^{n-1}}W\left(y_{(k-1)l^{n-1}+k'}\left|\left(u_{(h-1)l+1}^{hl}G_{\ast,k}\right)_{h=1}^{l^{n-1}}(G_{n-1})_{\ast,k'}\right.\right)\\
	\stackrel{(\ast)}{=}&\frac{1}{q^{l-1}}\sum_{u_{(i-1)l+j+1}^{il}}\prod_{k=1}^l\left[\frac{1}{q^{l^{n-1}-1}}\sum_{v_{i+1}^{l^{n-1}}}\prod_{k'=1}^{l^{n-1}}W\left(y_{(k-1)l^{n-1}+k'}\left|((u_{(h-1)l+1}^{hl}G_{\ast,k})_{h=1}^{i},v_{i+1}^{l^{n-1}})(G_{n-1})_{\ast,k'}\right.\right)\right]\\
	=&\frac{1}{q^{l-1}}\sum_{u_{(i-1)l+j+1}^{il}}\prod_{k=1}^lW_{n-1}^{(i)}\left(f(y_1^{l^n},u_1^{l(i-1)})_k\left|u_{(i-1)l+1}^{il}G_{\ast,k}\right.\right)\\
	=&\left(W_{n-1}^{(i)}\right)_1^{(j)}\left(f(y_1^{l^n},u_1^{l(i-1)}),u_{(i-1)l+1}^{(i-1)l+j-1}\left|u_{(i-1)l+j}\right.\right)
	\end{align*}
where  the equality $(\ast)$ follows from the fact that the space generated by the last  $l^n-(i-1)l-j+1$ rows in the  matrix $G_n$ has the same dimension as the space  $l$ times cartesian product of the space given by the   last $l^{n-1}-i+1$ rows in $G_{n-1}$. Therefore, since there is a bijection between the output alphabets of both channels, their parameters are the same.
\end{IEEEproof}
We will be interested in those channels where we can  \textit{recognize} the operations among their elements, more formally
\begin{definition} 
Let  $W:\mathbb{F}_q\rightarrow\mathcal{O}$ be a DMC. We say that $W$ is symmetric w.r.t the field addition if for each $x\in\mathbb{F}_q$ there is a permutation  $\sigma_x\in\mathrm{SG}(\mathcal{O})$ such that 
$$W(y|x)=W(\sigma_{x'-x}(y)|x'),\ \ \ \ \ \forall x,x'\in\mathbb{F}_Q,\ \forall y\in\mathcal{O}.$$
We say that $W$ is symmetric w.r.t. the field product if for each  $\alpha\in\mathbb{F}_q^\ast$ there is a permutation  $\psi_\alpha$ such that 
$$W(y|x)=W(\psi_\alpha(y)|\alpha x),\ \ \ \ \ \forall x\in\mathbb{F}_q,\ \forall y\in\mathcal{O}$$
We say that $W$ is symmetric w.r.t. the field operations in $\mathbb{F}_q$ (SOF) if $W$ is symmetric w.r.t. the field addition and product.
\end{definition}

\begin{remark} Note that if the channel
  $W$  is symmetric w.r.t the field addition  for each $\alpha\in\mathbb{F}_q$  we have that
$$W(y|x)=W(\sigma_\alpha(y)|\alpha+x).$$
\end{remark}

\begin{ex} Consider the channel  $W_{Sq}:\mathbb{F}_q\rightarrow\mathbb{F}_q$ with transition probabilities given by 
 $$W_{Sq}(y|x)=(1-p)\chi_{x}(y)+\frac{p}{n}.$$
Then 
$W_{Sq}$ is a  SOF channel. This channel has been studied in   \cite{qSC4,qSC3,qSC2,qSC1}.
\end{ex}

\begin{corollary}
	In the binary case  $q=2$ we have that any channel symmetric w.r.t the field addition is also a SOF channel.
\end{corollary}
\noindent Using Proposition~\ref{degchanpart} the polarization process can be analyzed inductively using the following result.
\begin{proposition}
If $W:\mathbb{F}_q\rightarrow\mathcal{O}$ is a SOF channel and  $G$ a non-singular square matrix of size $l\times l$ over   $\mathbb{F}_q$, then $W_1^{(i)}$ is also a SOF channel.
\end{proposition}
\begin{IEEEproof} The symmetry w.r.t the addition is known, see \cite{moriq}. Therefore we will check the  symmetry w.r.t the product.
Let $i\in\{1,\ldots,N\}$ and $\alpha\in\mathbb{F}_q^\ast$, then we have that
	\begin{align*}
	W_n^{(i)}(y_1^l,u_1^{i-1}|u_i)&=\sum_{u_{i+1}^N}W_n(y_1^N|u_1^N)\\
	&=\sum_{u_{i+1}^N}\prod_{k=1}^N W\left(y_k\left|\sum_{j=1}^N u_j(G_n)_{j,k}\right.\right)\\
	&=\sum_{u_{i+1}^N}\prod_{k=1}^N W\left(\psi_{\alpha}(y_k)\left|\alpha\sum_{j=1}^N u_j(G_n)_{j,k}\right.\right)\\
	&=W_n^{(i)}\left((\psi_{\alpha}(y_k))_{k=1}^N,\alpha u_1^{i-1} | \alpha u_i\right)
	\end{align*}
since $(u_{i+1},\ldots,u_N)\mapsto(\alpha u_{i+1},\ldots,\alpha u_N)$ is a bijection. Hence we define
$$\Psi_{\alpha}(y_1^N,u_1^{i-1})=((\psi_{\alpha}(y_k))_{k=1}^N,\alpha u_1^{i-1})$$
 and we get the result.  
\end{IEEEproof}
The following result also follows from \cite{moriq}.

\begin{proposition}\label{scsf}
Let  $G$ be a non-singular square matrix of size $l\times l$ over   $\mathbb{F}_q$, $V$  be an upper triangular invertible matrix and    $P$ a permutation matrix and consider $G'=VGP$. Let $W:\mathbb{F}_q\rightarrow\mathcal{O}$ a SOF channel and  $W_1^{(i)}$ y ${W'}_1^{(i)}$ the channels associated to the polarization processes with the matrices $G$ and $G'$ respectively. then we have
	$$I(W_1^{(i)})=I({W'}_1^{(i)}),$$
	$$Z(W_1^{(i)})=Z({W'}_1^{(i)}).$$
\end{proposition}

\begin{corollary}\label{mcfs}
	If  $G$ polarizes a SOF channel  $W$ and the matrices  $G'$ and  $G$ are given as in the above proposition, then   $G'$ also polarizes  $W$. Moreover,  if $\mathcal{A}_n$ and $\mathcal{A}'_n$ are the information sets generated by  $G$ and  $G'$ respectively, then
	$$\mathcal{A}_n=\mathcal{A}'_n.$$
\end{corollary}

\begin{IEEEproof}
	It follows from Proposition~\ref{scsf}  and  Proposition~\ref{degchanpart}.
\end{IEEEproof}

As we have seen before when the channel  $W$ is symmetric w.r.t. the addition the kernel of the polar code has all the information in the spaces
$$\langle G_{l,\ast}\rangle\subset\ldots\subset\langle G_{1,\ast},\ldots,G_{l,\ast}\rangle.$$
It is natural to associate this structure with the derivative of an algebraic curve. Let $\mathcal{X}$ be and algebraic curve and  $D=\sum_{i=1}^l P_i$, where $P_i\neq P_j$ if $i\neq j$, and $P_i$ rational points  (places in $\mathcal{X}$ of degree 1) and let us suppose that there exist divisors $A_1,\ldots,A_l$ such that the support of  $A_i$ and $D$ are disjoint for each $i$ , $A_i\leq A_{i+1}$ and
\begin{equation}
\mathbb{F}_q=C(D,A_1)\subsetneq\ldots\subsetneq C(D,A_l)=\mathbb{F}_q^l.
\label{kerag}
\end{equation}
We consider now
 $f_1,\ldots,f_l$ functions such that  $\langle f_1,\ldots,f_i\rangle =C(D,A_i)$ and we build the evaluation matrix  $G$ given by $G_{i,\ast}=ev_D(f_i)$.

Pointed algebraic curves satisfy the above construction. If we are given a pointed curve $(\mathcal{X},Q)$ and  $D=\sum_{i=1}^l P_i$  where  $P_i$ are different rational points, let $H^\ast(Q)=\{m_1,\ldots,m_l\}$ and  $A_i=m_i Q$, we get the desired structure.

\begin{ex} 
Consider the field with 4 elements $\mathbb{F}_4$ and the Hermitian curve $x^3=y^2+y$. If we take  $Q$ as the common pole of $x$ and  $y$  and the divisor $D=\sum P_{\alpha,\beta}$ where $P_{\alpha,\beta}$ is the common zero of $x-\alpha$ and $y-\beta$, then $\deg D=8$ and  $H^\ast(Q)=\{0,2,3,4,5,6,7,9\}$. It follows that 
	
	$$\begin{array}{r|cccccccc}

& 00 & 01 & 1\alpha & 1\alpha^2& \alpha\alpha & \alpha\alpha^2 & \alpha^2\alpha & \alpha^2\alpha^2\\\hline

x^3y&0&0&\alpha&\alpha^2&\alpha&\alpha^2&\alpha&\alpha^2\\

x^2y & 0& 0&\alpha&\alpha^2&1&\alpha&\alpha^2&1\\

x^3& 0&0&1&1&1&1&1&1\\

xy & 0&0&\alpha&\alpha^2&\alpha^2&1&1&\alpha\\

x^2& 0&0&1&1&\alpha^2&\alpha^2&\alpha&\alpha\\

y&0&1&\alpha&\alpha^2&\alpha&\alpha^2&\alpha&\alpha^2\\

x&0&0&1&1&\alpha&\alpha&\alpha^2&\alpha^2\\

1&1&1&1&1&1&1&1&1
\end{array}$$
	
\end{ex}

When the channel   $W$  is symmetric w.r.t. the addition we will call \emph{the  kernel associated to the pointed curve $(\mathcal{X},Q)$} to any evaluation matrix generated by a basis $\{f_1,\ldots,f_l\}$ where each $f_i\in\mathcal{L}(m_iQ)\setminus\mathcal{L}(m_{i-1}Q)$. Note that it is well defined since by Corollary~\ref{mcfs} any matrix of this form produces the same set $\mathcal{A}_n$.

In order to study the structure of those matrices associated to curves note that   $\mathcal{L}(\infty Q)=\bigcup_{m=0}^\infty \mathcal{L}(mQ)$ is a finitely generated algebra.

\begin{proposition}[\cite{ruudorder}, Proposition 5.2]\label{polrayo}
	Let $(\mathcal{X},Q)$ be a pointed curve and $H(Q)=\langle a_1,\ldots,a_s\rangle$, where $\{a_1,\ldots,a_s\}$  is a minimal generator set of $H(Q)$, then there exists an ideal   $I\subset\mathbb{F}_q[t_1,\ldots,t_s]$ such that
	$$\mathcal{L}(\infty Q)=\mathbb{F}_q[t_1,\ldots,t_s]/I.$$
\end{proposition}

\begin{proposition}
Let $D=\sum_{i=1}^l P_i$ be a divisor of rational places and let us suppose that there exists a  $z\in\mathcal{L}(\infty Q)$ such that
$$(z)=D-lQ.$$
If $f_z\in\mathbb{F}_q[t_1,\ldots,t_s]$ is a polynomial such that $f_z$ represents $z$, then   
$$ev_D(\mathcal{L}(\infty Q))=\mathbb{F}_q[t_1,\ldots,t_s]/\langle I,f_z\rangle.$$
\end{proposition}

\begin{IEEEproof}
	If $x\in\ker(ev_D)\cap\mathcal{L}(mQ)$, then $x\in\mathcal{L}(mQ-D)$ and since $(z)=D-lQ$ then we have
	$$x\in z\mathcal{L}((m-l)Q).$$
	Ie., the image of  $x$ in $\mathbb{F}_q[t_1,\ldots,t_s]$ is in the ideal generated by the equivalence class represented by  $f_z$, hence $x\in\mathbb{F}_q[t_1,\ldots,t_s]/\langle I,f_z\rangle$. Since  $m$   has been arbitrary chosen we have $\subset$. The other contention follows  since $ev_D(yz)=ev_D(y)\ast ev_D(z)=0$.
\end{IEEEproof}

\section{Information sets for SOF channels}
In this section we analyze the information set $\mathcal{A}_n$ for a SOF channel. The main tool we will use is channel degradation.

\begin{definition}\normalfont
	Let  $W:\mathcal{I}\rightarrow\mathcal{O}$ and $W':\mathcal{I}\rightarrow\mathcal{O}'$ be two DMC channels. We say the $W'$ is a degradation of  $W$ and we will denote it as $W'\preceq W$, if there exists a channel   $W'':\mathcal{O}\rightarrow\mathcal{O}'$ such that
	$$W'(y|x)=\sum_{z\in\mathcal{O}}W''(y|z)W(z|x)$$
 for any $y\in\mathcal{O}',x\in\mathcal{I}$.
\end{definition}

One can think on degradation as a ``composition" of channels in the sense that the transition probability of $W'$ represents the probability of the event if we send   $x$ trough channel $W$ and the received is transmited by channel  $W''$ we get  $y$. That is

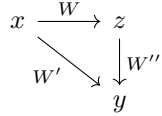
\[\begin{tikzcd}
x\arrow[r, "W"]\arrow{rd}{}[swap]{W'}  &z \arrow[d, "W''"]\\
& y
\end{tikzcd}\]
Therefore degradation make the transmission gets worse.

\begin{proposition}\label{degpar}
If channels  $W:\mathcal{I}\rightarrow\mathcal{O}$ and $W':\mathcal{I}\rightarrow\mathcal{O'}$ satisfy $W'\preceq W$, then  
	\begin{align*}
		Z(W)&\leq Z(W'),\\
		I(W)&\geq I(W').
	\end{align*}
\end{proposition}

\begin{IEEEproof}
Let $a,b\in\mathcal{I}$, then  
\begin{align*}
Z_{a,b}(W')&=\sum_{y\in\mathcal{O'}}\sqrt{W'(y|a)W'(y|b)}\\
&=\sum_{y\in\mathcal{O'}}\sqrt{\sum_{z\in\mathcal{O}} W''(y|z)W(z|a)\sum_{z\in\mathcal{O}} W''(y|z)W(z|b)}\\
&\geq \sum_{y\in\mathcal{O'}}\sum_{z\in\mathcal{O}}\sqrt{W(z|a)W(z|b)}W''(y|z)\\
&=\sum_{z\in\mathcal{O}}\sqrt{W(z|a)W(z|b)}\\
&=Z_{a,b}(W)
\end{align*}
\noindent where the inequality follows from  Cauchy-Schwartz. If we take the mean among all the pairs   $(a,b)\in\mathcal{I}^2$, $a\neq b$ it follows the desired result.
The second inequality follows form the data processing inequality \cite{dpi}.
\end{IEEEproof}

Moreover, degradation is preserved by the polarization process, more formally  

\begin{proposition}\label{degpolar}
Let $W:\mathbb{F}_q\rightarrow\mathcal{O}$ and $W':\mathbb{F}_q\rightarrow\mathcal{O}'$ be two channels such that $W'\preceq W$ and let $G$  be a non-singular square matrix  of size $l\times l$ over the finite field  $\mathbb{F}_q$, then 
	$${W'}_1^{(i)}\preceq W_1^{(i)}.$$
\end{proposition}

\begin{IEEEproof}
\begin{align*}
		{W'}_1^{(i)}(y_1^l,u_1^{i-1}|u_i)&=\sum_{u_{i+1}^l}\prod_{k=1}^l W'(y_k|uG_{\ast,k})\\
		&=\sum_{u_{i+1}^l}\prod_{k=1}^l\sum_{z\in\mathcal{O}} W''(y_k|z)W(z|uG_{\ast,k})\\
		&=\sum_{u_{i+1}^l} \sum_{z_1^l}\prod_{k=1}^l W''(y_k|z_k)W(z_k|uG_{\ast,k})\\
		&=\sum_{z_1^l}\prod_{k=1}^l W''(y_k|z_k)\sum_{u_{i+1}^l}\prod_{k=1}^lW(z_k|uG_{\ast,k})\\
		&=\sum_{z_1^l}\prod_{k=1}^l W''(y_k|z_k)W_1^{(i)}(z_1^l,u_1^{i-1}|u_i).
	\end{align*}
If we define $W'''(y_1^l,u_1^{i-1}|z_1^l,u_1^{i-1})=\prod_{k=1}^l W''(y_k|z_k)$  we conclude the proof. 
\end{IEEEproof}

\begin{lemma}\label{lemma}
	Let $(\mathcal{X},Q)$ be a pointed curve of genus $g$ such that  $l\geq 2g$ and $H(Q)=\langle a_1,\ldots,a_s\rangle$ is a minimal generator set of $H(Q)$. Let us define $f_i:H^\ast(Q)\rightarrow H^\ast(Q)$ as
	$$f_i(m)=\begin{cases}
		m-a_i& m-a_i\in H^\ast(Q)\\
		l+m-a_i& m-a_i\notin H^\ast(Q)
		\end{cases},$$
then $f_i$ is a bijection.
\end{lemma}

\begin{IEEEproof}
	Since $l\geq 2g$ we know that $H^\ast(Q)=H(Q)\setminus\{l+H(Q)\}=H(Q)\cap \{0,\ldots,l\}\cup\{l+l_1,\ldots,l+l_g\}$, where $l_i$ are the gaps of $Q$.
	
If $a_i\leq m<n$, then either $m-a_i\in H(Q)$ (therefore in $H^\ast(Q)$) or $m-a_i\notin H(Q)$ and hence $l+m-a_i\in H^\ast(Q)$, while $l+m\notin H^\ast(Q)$ (if not  $m$ will be a  gap).
	
	If $m>l$ and $m-a_i<l$, then   $m-a_i\in H^\ast(Q)$ but $m-a_i\notin f_i(\{a_i,\ldots,l-1\}\cap H(Q))\subset \{0,\ldots,l-a_i-1\}$.
	
	On the other hand, if $m>l$ y $m-a_i>l$  then $m-a_i\in H^\ast(Q)$.  Of course, if $m-a_i\notin H^\ast(Q)$ then $m-a_i-l$ is a non-gap, but $m-l$ is a gap, which contradicts $a_i\in H(Q)$. Therefore  $m-a_i\neq f_i(m')$ for any $a_i\leq m'<l$. 
	
	Finally, if $m<a_i$ then $-a_i\leq m-a_i<0$ and hence $l-a_i\leq l+m-a_i<l$ and since $m$ is a non-gap it is not covered in the previous cases, therefore  $f_i$ is injective and by cardinality it is bijective. 
\end{IEEEproof}

From now on  $T=(t_1,\ldots,t_s)$ y sea $R[T]=\mathbb{F}_q[t_1,\ldots,t_s]/I(T)$, where the ideal $I(T)$ is the one given in Proposition~\ref{polrayo}.

\begin{theorem}\label{degsem}
	Let $W$ be a SOF channel. Let us consider the pointed curve $(\mathcal{X},Q)$ and a divisor on the curve $D=(z)_0$ with $l=\deg (z)_0\geq 2g$, and let  $H^\ast(Q)=\{m_0,\ldots,m_{l-1}\}$. If we consider also an element $m_{l-i}\in H^\ast(Q)$ with $m_{l-i}<l$. If $m_{l-j}=m_{l-i}-a_r\in H^\ast(Q)$, where $a_r$ is one of the generators of $H(Q)$, then
	$$W_1^{(i)}\preceq W_1^{(j)}.$$
\end{theorem}

\begin{IEEEproof}
We can choose monomials $M_k$ in $R[T]$ such that $v_Q(M_k)=m_{l-k}$ and such that  if $m_{l-k}-a_r\in H^\ast(Q)$ then $t_r|M_k$ and also if $b=\max\{a\in\mathbb{N}_0\ |\ t_r^a|f,\ f\in\mathcal{L}(m_{r-k}Q)\setminus\mathcal{L}((m_{r-k}-1)Q)\}$ then $t_r^b|M_k$.  We construct the kernel  $G$ evaluating those monomials. Thus if $t_r|M_k$ for any $k$ then $\frac{M_k}{t_r}=M_{k'}$ for some $k'$ it is clear that $m_{l-k}=m_{l-k'}+a_r$.
	
Consider the polynomial $f=\sum_{k=1}^l u_kM_k$ and denote as $W_1(y|f)=\prod_{k=1}^l W(y_k|f(P_{k}))$. We have that  $W_1(y|f)=W_1(y|u_1^l)$. If we consider $A=\{k\in\{1,\ldots,l\}|\ m_{l-k}-a_r\in H^\ast(Q)\}$, then
	$$W_1(y|f)=W_1\left(y\left|\sum_{k\in A} u_kM_k+\sum_{k\notin A} u_kM_k\right.\right).$$ 

	Let $F_r$ the function  $F_r(M_k)=M_{k'}\Leftrightarrow f_r(m_{l-k})=m_{l-k'}$. Applying Lemma~\ref{lemma} we have a bijection of the chosen monomials and also  
	$$W_1(y|f)=W_1\left(y\left|\sum_{k\in A}u_kF_i(M_k)t_r+\sum_{k\notin A} u_kM_k\right.\right),$$
	where  $u_iM_i=u_iM_jt_r$. We define $\overline{y}=(y_{\alpha_1},\ldots,y_{\alpha_z})$ where $\mathrm{supp}\ t_r=\{\alpha_1<\ldots<\alpha_z\}$ and if  $g$ is a polynomial we define
	$$\sigma_g(\overline{y})=(\sigma_{g(P_{\alpha_1})}(y_{\alpha_1}),\ldots,\sigma_{g(P_{\alpha_z})}(y_{\alpha_z})),$$
	$$\psi_{t_r^{-1}}(\overline{y})=\left(\psi_{t_r^{-1}(P_{\alpha_1})}(y_{\alpha_1}),\ldots,\psi_{t_r^{-1}(P_{\alpha_z})}(y_{\alpha_z})\right).$$
Note that 
	{
	\begin{align*}
		W_1(y|f)&=W_1\left(y\left|\sum_{k\in A} u_kF_r(M_k)t_r+\sum_{k\notin A}u_kM_k\right.\right)\\
		&=\prod_{\alpha\notin\mathrm{supp}\ t_r}W\left(y_\alpha\left|\sum_{k\notin A}u_kM_k(P_{\alpha})\right.\right)\prod_{h=1}^zW\left(y_{\alpha_h}\left|\left(\sum_{k\in A}u_kF_r(M_k)t_r+\sum_{k\notin A}u_kM_k\right)(P_{\alpha_h})\right.\right)
	\end{align*}
	}
	Let  $\hat{u}$ be the result of taking only those indexes  $u_k$ with $k\notin A$ and $g(\hat{u})=\displaystyle\sum_{k\notin A}u_k(F_r(M_k)t_r-M_k)$. Since we are in a SOF channel it follows
	{ 
	\begin{align*}
		W_1(y|f)&=\prod_{\alpha\notin\mathrm{supp}\ t_r}W\left(y_\alpha\left|\sum_{k\notin A}u_kM_k(P_{\alpha})\right.\right)\prod_{h=1}^zW\left(y_{\alpha_h}\left|\left(\sum_{k\in A}u_kF_r(M_k)t_r+\sum_{k\notin A}u_kM_k\right)(P_{\alpha_h})\right.\right)\\
		&=\prod_{\alpha\notin\mathrm{supp}\ t_r}W\left(y_\alpha\left|\sum_{k\notin A}u_kM_k(P_{\alpha})\right.\right)\prod_{h=1}^zW\left(\sigma_{g(\hat{u})}(\overline{y})_h\left|\left(t_r\sum_{k=1}^l u_kF_i(M_k)\right)(P_{\alpha_h})\right.\right)\\
		&=\prod_{\alpha\notin\mathrm{supp}\ t_r}W\left(y_\alpha\left|\sum_{k\notin A}u_kM_k(P_{\alpha})\right.\right)\prod_{h=1}^zW\left(\psi_{t_r^{-1}}\left(\sigma_{g(\hat{u})}(\overline{y})\right)_h\left|\sum_{k=1}^l u_kF_r(M_k)(P_{\alpha_h})\right.\right).
		\end{align*}}
Now since  $F_r$ is a bijection there is a permutation   $\varphi:\{1,\ldots,l\}\rightarrow\{1,\ldots,l\}$ such that
	$$\sum_{k=1}^l u_kF_r(M_k)=\sum_{k=1}^lu_{\varphi(k)}M_k,$$
that also satisfies
$$\varphi(j)=i$$
$$\varphi^{-1}(k)>j\Longleftrightarrow k\in A,\ k>i.$$
This last fact is because if $m_{l-k}-a_r\notin H^\ast(Q)$ then $l+m_{l-k}-a_r\geq l-a_r>m_{l-i}-a_r$. Now let us
consider the channel given by $Q:\mathcal{Y}^l\times\mathbb{F}_q^{j-1}\rightarrow\mathcal{Y}^l\times\mathbb{F}^{i-1}$ as follows
$$Q(y,u_1^{i-1}|z,v_1^{j-1})=\prod_{\alpha\notin\mathrm{supp}\ t_r}W\left(y_\alpha\left|\sum_{k\notin A} v_{\varphi^{-1}(k)}M_{k}(P_{\alpha})\right.\right)$$
if $u_{\varphi(k)}=v_k$ for $1\leq k\leq i-1$ and  $\overline{y}=\sigma_{g(v)}\left(\psi_{t_r}(\overline{z})\right)$ with $g(v)=\sum_{k\notin A} v_{\varphi^{-1}(k)}(F_r(M_k)-M_k)$ and $0$ elsewhere. 

\noindent If $v\in\mathbb{F}_q^l$ is a vector where $v_j=u_i$, then
\begin{align*}
&\sum_{z,v_1^{j-1}}Q(y,u_1^{i-1}|z,v_1^{j-1})W_1^{(j)}(z,v_1^{j-1}|u_i)\\\stackrel{(\ast)}{=}&\frac{1}{q^{l-1}}\sum_{z,v_1^{j-1}}\sum_{v_{j+1}^l}\prod_{\alpha\notin\mathrm{supp}\ t_r}W\left(y_\alpha\left|\sum_{k\notin A}v_{\varphi^{-1}(k)}M_k(P_{\alpha})\right.\right)W_1(z|v)\\
=&\frac{1}{q^{l-1}}\sum_{z,v_1^{j-1}}\sum_{v_{j+1}^l}\prod_{\alpha\notin\mathrm{supp}\ t_r}W\left(y_\alpha\left|\sum_{k\notin A}v_{\varphi^{-1}(k)}M_k(P_{\alpha})\right.\right)\prod_{\alpha\notin\mathrm{supp}\ t_r}W\left(z_\alpha\left|\sum_{k\notin A}v_kM_k(P_{\alpha})\right.\right)\\
&\prod_{h=1}^zW\left(\psi_{t_r^{-1}}\left(\sigma_{g(v)}(\overline{y})\right)_h\left|\sum_{k=1}^l v_kM_k(P_{\alpha_h})\right.\right)\\
=&\frac{1}{q^{l-1}}\sum_{v_1^{j-1}}\sum_{v_{j+1}^l}\prod_{\alpha\notin\mathrm{supp}\ t_r}W\left(y_\alpha\left|\sum_{k\notin A}v_{\varphi^{-1}(k)}M_k(P_{\alpha})\right.\right)\prod_{h=1}^zW\left(\psi_{t_r^{-1}}\left(\sigma_{g(v)}(\overline{y})\right)_h\left|\sum_{k=1}^l v_kM_k(P_{\alpha_h})\right.\right)
\end{align*}
\begin{align*}
=&\frac{1}{q^{l-1}}\sum_{u_{i+1}^l}\prod_{\alpha\notin\mathrm{supp} t_r}W\left(y_\alpha\left|\sum_{k\notin A}u_kM_k(P_{\alpha})\right.\right)\prod_{h=1}^zW\left(\psi_{t_r^{-1}}\left(\sigma_{g(\hat{u})}(\overline{y})\right)_h\left|\sum_{k=1}^lu_{\varphi(k)}M_k(P_{\alpha_h})\right.\right)\\
=&\frac{1}{q^{l-1}}\sum_{u_{i+1}^l}W_1(y|u)\\
=&W_1^{(i)}(y,u_1^{i-1}|u_i)
\end{align*}

\noindent where from step $(\ast)$ on the sum that ranges in $z,v_1^{j-1}$ is only over those indexes that $\overline{z}=\psi_{t_r^{-1}}(\sigma_{g(v)}(\overline{y}))$ and $u_{\varphi(k)}=v_k$ for $1\leq k\leq i-1$. Finally since for any matrix 
$G$ defining the kernel the information set does not change then the result follows. 
\end{IEEEproof}

In order to fix a matrix given a pointed curve $(\mathcal{X},Q)$ so we can describe the polar code in terms of the function field we will take  $D=(z)_0=\sum_{i=0}^{l-1} P_i$ and $H^\ast(Q)=\{m_1=0,\ldots,m_l\}$. It is known that $ev_D(\mathcal{L}(\infty Q))=R[T]/\langle f_z\rangle$, thus a basis is given by $ev_D(\mathcal{L}(\infty Q))=\Delta(I(T),f_z)=\{M_0,\ldots,M_{l-1}\}$ where $(M_i)_\infty=m_{i+1}Q$.  Evaluating that basis we will construct the matrix $G$ for the polarization process. From now on we shall consider always the matrix for constructing polar codes from the pointed curve  $(\mathcal{X},Q)$.

For each  $n\in\mathbb{N}$ and each $i\in\{0,\ldots,l^n-1\}$ we denote by $(i_n,\ldots,i_1)$ to the  $l$-ary expansion of $i$, ie.  $0\leq i_k\leq l-1$ and
$$i=\sum_{k=1}^n i_kl^{k-1}.$$
Let  $P^n_i=(P_{i_n},\ldots,P_{i_1})$, $\mathbb{F}_q[X_1,\ldots,X_n]$, $X_k=(x_{k1},\ldots,x_{ks})$ and
$$I_n=\langle I(X_1),\ldots,I(X_n),f_z(X_1),\ldots,f_z(X_n)\rangle.$$
In the polynomial ring  $\mathbb{F}_q[X_1,\ldots,X_n]$ we take the monomial ordering inherit from the weights in the variables   $x_i$, ie. the monomial ordering defined by the vectors with entries 
$$w_{n+1-i,il+j}=-v_Q(x_{ij}),$$
and we will break ties with RevLex if it is needed. As a resume, we get $n$ copies of the original ring $R[T]$ and order it with weights inherit from the valuation in  $Q$.

\begin{proposition}\label{orderg}
	Let $M^n_k(X_1,\ldots,X_n)=M_{k_{1}}(X_1)\cdots M_{k_n}(X_n)$, where $(k_n,\ldots,k_1)$ is $l$-ary expansion of $0\leq k\leq l^n-1$ and 
	$$M^n_k(P^n_j)=M_{k_1}(P_{j_n})\cdots M_{k_n}(P_{j_1}),$$ 
then $\Delta(I_n)=\{ M^n_k(X_1,\ldots,X_n)\ |\ k\in\{0,\ldots,l^n-1\}$ and the matrix $G_n$ in the polarization process satisfies
 	$$G_n(i,j)=M^n_{(l^n-i)}(P^n_j).$$
\end{proposition}

\begin{IEEEproof}
	The equality in $\Delta(I_n)$ is clear. The proof of the statements related with   $G_n$ and the columns are also clear since it is just an application of the Kronecker product. For checking the property on  the rows we will use induction.
	 Case $n=1$ is clear so let us suppose it is true in the step $n-1$.
	
\noindent Due to the bit-reversal in the polarization matrix we know that the row   $jl^{n-1}+i$whose  $l$-ary expansion is $(j,i_n,\ldots,i_1)$ is the row in the Kronecker product's matrix with  $l$-ary expansion $(i_1,\ldots,i_n,j)$. That row is in correspondence with the product of the monomials   $M_j(X_n)$ and $M^n_{l^{n-1}-i}(X_1,\ldots,X_{n-1})$ by the induction hypothesis, so the result follows. 
\end{IEEEproof} 
\noindent As a corollary of
 Theorem~\ref{degsem} we have
\begin{corollary}\label{order1}
	Let $G$ be the matrix associated to the pointed curve $(\mathcal{X},Q)$ and $M_i\in\Delta(I(X_1),f_z)$ with $\deg M_i\leq l$. If $M_i\neq\Delta(I,M_j)$ with $i>j$, then
	$$W_1^{(l-i)}\preceq W_1^{(l-j)}$$
In particular the result follows if $M_j|M_i$ (where the division is in the ring $R[T]$).
\end{corollary}
Consider the set $\mathcal{A}_n\subset\Delta(I_n)$ and the code given by
$$C_{\mathcal{A}_n}=\langle\{(M^n_k(P^n_j))_{j=0}^{l^n-1}\ |\ M^n_k\in\mathcal{A}_n\}\rangle,$$
we say that $C_{\mathcal{A}_n}$ is a polar code if for each $M^n_k\in\mathcal{A}_n$ and for each $M^n_j\notin\mathcal{A}_n$ we have that
$$Z(M^n_k):=Z(W_n^{(l^n-k)})\leq Z(M^n_j).$$

\begin{proposition}\label{infsetd}
	Let $C_{\mathcal{A}_n}$ be a polar code constructed from a pointed curve  $(\mathcal{X},Q)$. If $M^n_i\in\mathcal{A}_n$ satisfies for all $i_k<j_k$, $\deg M_{i_k}\leq l$ and $M_{i_k}\notin\Delta(I(X_k),M_{j_k})$, $1\leq k\leq n$, then $M^n_j\in\mathcal{A}_n$.
	In particular, if $M^n_j|M^n_i$ then $M^n_j\in\mathcal{A}_n$.
\end{proposition}

\begin{IEEEproof}
It follows from induction taking into account that
	 $$W\preceq W'\Longrightarrow W_1^{(k)}\preceq {W'_1}^{(k)}$$
and
	 $$\left(W_{n-1}^{(k)}\right)^{(k')}=W_n^{((k-1)l+k')}.$$
Thus if $i_1<j_1$, $M^n_i=M_{i_1}(X_1)M^{n-1}_{i'}$ and $M^n_j=M_{j_1}(X_1)M^{n-1}_{j'}$ then
\begin{align*}
W_n^{(l^n-i)}&=W_n^{((l^{n-1}-i'-1)l+l-i_1)}\\&=(W_{n-1}^{(l^{n-1}-i')})^{(l-i_1)}\\&
\preceq (W_{n-1}^{(l^{n-1}-i')})^{(l-j_1)}\\&\preceq (W_{n-1}^{(l^{n-1}-j')})^{(l-j_1)}\\&=W_n^{(l^n-j)}
\end{align*}
The induction step follows from Corollary~\ref{order1} above.
\end{IEEEproof}

\begin{definition}
	We say that the code $C_{\mathcal{A}_n}$ is weakly decreasing if for all  $M^n_k\in\mathcal{A}_n$ we have that if $j$ satisfies $k_i\geq j_i$, $(j_n,\ldots,j_1)$ (the $l$-ary expansion of $j$) then $M^n_j\in\mathcal{A}_n$. 
\end{definition}

\begin{remark} The name \emph{weakly decreasing} is recovered from the one in 
	 \cite{Bardet}. We do not have a way of ensuring that a code is weakly decreasing, but from the fact that any polar code is the shortening of a weakly decreasing code, using the proposition above we will check that for some cases the difference between a polar code and weakly decreasing code is not so big (measured as the number of rows that one has to remove).
\end{remark}

\begin{ex} Consider the hermitian curve  $x^3=y^2+y$ over  $\mathbb{F}_4$ pointed in $Q$ the common pole of  $x$ and $y$. 
In this case $$\Delta(I,x^4-x)=\{x^3y,x^2y,x^3,xy,x^2,y,x,1\}$$ that correspond with the values  $H^\ast(Q)=\{9,7,6,5,4,3,2,0\}$. If we choose   $x^2y\in\mathcal{A}_1$ then $xy,x^2,y,x,1\in\mathcal{A}_1$; if $x^3\in\mathcal{A}_1$ then we have a weakly decreasing code. On the other hand if $x^3\in\mathcal{A}_1$ then $x^2,y,x,1\in\mathcal{A}_1$ and for getting a weakly decreasing code is enough to see that  $xy\in\mathcal{A}_n$. 
\end{ex}
	
\begin{corollary}
	Rational curves provide kernels for  polar codes that are weakly decreasing.
\end{corollary}

\begin{IEEEproof}
	Those curves there are no  gaps and  $H^\ast(Q)=\{0,1,\ldots,q-1\}$ therefore for all  $m\in H^\ast(Q)$, $m<n$ and $m-1\in H^\ast(Q)$.
\end{IEEEproof}

\begin{remark} The result stated in the corollary above generalizes 
	the same statement \cite{Bardet} for rational curves over  $\mathbb{F}_2$.
\end{remark}

\begin{definition} We say that a code 
	 $C_{\mathcal{A}_n}$  is decreasing if it is weakly decreasing and there are $h_1,\ldots,h_k\in\{0,\ldots,l-1\}$ (maybe not distinct) such that
	$$M_{h_1}(X_{i_1})\cdots M_{h_k}(X_{i_k})\in\mathcal{A}_n,$$
	then for any  $j_v\leq i_v$, $v\in\{1,\ldots,k\}$ we have
	$$M_{h_1}(X_{j_1})\cdots M_{h_k}(X_{j_k})\in\mathcal{A}_n.$$
\end{definition}
This extra property for being decreasing will be called degrading property. Indeed it makes sense since in each step in the polarization process the new elements are worse than the previous ones.

\begin{proposition}\label{polardegr}
	If $C_{\mathcal{A}_n}$  is a polar code and $M_{h_1}(X_{i_1})\cdots M_{h_k}(X_{i_k})\in\mathcal{A}_n$ then
	$$M_{h_1}(X_{j_1})\cdots M_{h_k}(X_{j_k})\in\mathcal{A}_n$$
	for all  $j_v\leq i_v$, $1\leq v\leq k$.
\end{proposition}

\begin{IEEEproof}
	We will prove it by induction on  $n$. For  $n=2$ remember that  $G_2=B_2G^{\otimes 2}$ where $B_2$ interchanges the $i$-th row  with  $l$-ary expansion  $(i_2,i_1)$ with the one with expansion $(i_1,i_2)$ (rows are indexed from $0$ to $l^2-1$) and therefore  $(B_2)^{-1}=B_2$. Moreover $G_2B_2=G^{\otimes 2}$. We have that
	$${G_2}_{i,j}=M_{i_1}(P_{j_2})M_{i_2}(P_{j_1})$$
	that multiplied by  $B_2$  returns
	$${G_2B_2}_{i,j}=M_{i_1}(P_{j_1})M_{i_2}(P_{j_2})={G^{\otimes 2}}_{i,j}.$$
	
Hence if $u\in\mathbb{F}_q^{l^2}$, then
	\begin{equation}
	(uB_2)(G_2B_2)=u(G^{\otimes_2}B_2)=uG_2\tag{$\ast$}.
	\end{equation}
	Moreover note that the last  $l$ entries in $uB_2$ correspond with  $(u_{l-1},u_{2l-1},\ldots,u_{l^2-1})$. Whence suppose that   $M_h(X_2)\in\mathcal{A}_2$; that monomial is associated with the row  $l^2-lh=l(l-h)$ while the monomial $M_h(X_1)$ is associated with $l^2-h$. If we define $Q:\mathcal{Y}^{l^2}\times\mathbb{F}_q^{l^2-h-1}\rightarrow\mathcal{Y}^{l^2}\times\mathbb{F}_q^{l^2-lh-1}$ with probabilities
	$$Q(y_1^{l^2},u_1^{l^2-lh-1}|z_1^{l^2},v_1^{l^2-h-1})=1$$
	if  $B_2y_1^{l^2}=z_1^{l^2}$ and $(vB_2)_1^{l^2-lh-1}=u_1^{l^2-lh-1}$.Then  by  $(\ast)$, it follows that
\begin{align*}
&\sum_{z_1^{l^2},v_{1}^{l^2-h-1}}Q(y_1^{l^2},u_1^{l^2-lh-1}|z_1^{l^2}v_1^{l^2-h-1})W_2^{l^2-h}(z_1^{l^2},v_1^{l^2-h-1}|u_{l^2-lh})\\
=&\sum_{z_1^{l^2},v_{1}^{l^2-h-1}}\sum_{v_{l^2-h+1}^{l^2}}\frac{1}{q^{l^2-1}}Q(y_1^{l^2},u_1^{l^2-lh-1}|z_1^{l^2}v_1^{l^2-h-1})W_2(z|v)\\
=&\sum_{u_{l^2-lh+1}^{l^2}} \frac{1}{q^{l^2-1}}W_2(B_2y|uB_2)\\
=&\sum_{u_{l^2-lh+1}^{l^2} }\frac{1}{q^{l^2-1}}W_2(y|u)\\
=&W_2^{(l^2-lh)}(y_1^{l^2},u_1^{l^2-lh-1}|u_{l^2-lh}).
\end{align*}
In other words, $W_2^{l^2-lh}\preceq W_2^{l^2-h}$  and therefore $M_h(X_1)\in\mathcal{A}_2$.

Let us suppose that it is true for  $n-1$.
Note that  $\left(W_{n-1}^{(i)}\right)^{(l)}=W_n^{(li)}$, hence if $M_{h_1}(X_{i_1})\cdots M_{h_k}(X_{i_k})\in\mathcal{A}_n$ and $i_1\geq 2$ we have that  $M_{h_1}(X_{j_1})\cdots M_{h_k}(X_{j_k})\in\mathcal{A}_n$ for all $j_v\leq i_v$, $v\in\{1,\ldots,k\}$ such that  $2\geq j_1\leq i_1$.  If  $2\geq j_1<i_1$ then we have that if  $i=\sum_{v=1}^k h_vl^{i_v-1}$ and $j=\sum_{v=1}^k h_vl^{j_v-1}$ then $l|i$ and $l|j$ by the induction hypothesis in $n-1$, therefore  
$$W_n^{(l^n-i)}=\left(W_{n-1}^{\left(\frac{l^n-i}{l}\right)}\right)^{(l)}\preceq\left(W_{n-1}^{\left(\frac{l^n-j}{l}\right)}\right)^{(l)}=W_n^{(l^n-j)},$$
and the result follows. Same reasoning guaranties the result if  $i_1=j_1=1$. It only remains the case $1=j_1<i_1$. Taking into account that degradation is a transitive relation we can suppose that  $i_1=2$ and choosing  $i'=i-h_1l$ an $j'=j-h_1$, where $i$ and $j$ are as before. Thus applying the induction step to the case $n-2$, we have that  $W_{n-2}^{\left(\frac{l^n-i'}{l^2}\right)}\preceq W_{n-2}^{\left(\frac{l^n-j'}{l^2}\right)}$. Applying induction for the case   $2$ 
we have
\begin{align*}
W_n^{(l^n-i)}&=\left(W_{n-2}^{\left(\frac{l^n-i'}{l^2}\right)}\right)_2^{(l^2-lh_1)}\\
&\preceq\left(W_{n-2}^{\left(\frac{l^n-i'}{l^2}\right)}\right)_2^{(l^2-h_1)}\\
&\preceq\left(W_{n-2}^{\left(\frac{l^n-j'}{l^2}\right)}\right)_2^{(l^2-h_1)}\\
&=W_n^{(l^n-j)}
\end{align*}
and we conclude the proof.
\end{IEEEproof}
	
\begin{remark}The previous result does not make  use of the property SOF. Thus a polar code weakly decreasing is decreasing.
\end{remark}

\begin{ex}\label{hermit2}\normalfont
	Take the hermitian curve from previous examples and $n=2$. If $y_1x_2\in\mathcal{A}_2$ then, using Proposition~\ref{infsetd} we get
	$$x_2,y_1,1\in\mathcal{A}_2$$
	and applying the proposition above $x_1\in\mathcal{A}_2$. If $x_1x_2\in\mathcal{A}_2$, with the previous elements we have a descending polar  code.
\end{ex}

\section{Minimum Distance and Dual of Polar Codes}

Let us check some properties of the structure of a polar code constructed from a pointed curve $(\mathcal{X},Q)$.

\begin{proposition}\label{minpol}
	Let $C_{\mathcal{A}_n}$ be a decreasing code and let $(K_n,\ldots,K_1)$ be a tuple such for each $M^n_k\in\mathcal{A}_n$ the $l$-ary expansion of $k$, $(k_n,\ldots,k_1)$, satisfies $k_i\leq K_i$ for each $i\in\{1,\ldots,n\}$. Take $H^\ast(Q)=\{m_1,\ldots,m_l\}$ and $d_i=\delta(C(\mathcal{X},D,m_{K_i+1}Q))$ and let $k'$ be such that $M^n_k\in\mathcal{A}_n$ for each $k'\geq k$ and $d'_i=\delta(C(\mathcal{X},D,m_{k'_i+1}Q))$, then we have
	$$\prod_{i=1}^n d'_i\geq \delta(C_{\mathcal{A}_n})\geq\prod_{i=1}^n d_i.$$
\end{proposition}

\begin{IEEEproof}
	We will proceed by induction. It is clear for $n=1$. Let us suppose it is true for $n$ and get the result for $n+1$. First note  that $K_1\geq K_2\geq\ldots\geq K_{n+1}$ since if $M_k^n\in\mathcal{A}_n$ and $M_{K_j}(X_j)|M_k^n$ (from the hypothesis in the proposition), then $M_{K_j}(X_j)\in\mathcal{A}_n$ and $M_{K_j}(X_{j-1})\in\mathcal{A}_n$, since the code is  decreasing and therefore $K_j\leq K_{j-1}$. Let $C_1$ be the generator matrix of $C(\mathcal{X},D,m_{K_1+1}Q)$ and let $A$ be the matrix with rows the evaluations of $M_k^{n+1}\in\mathcal{A}_{n+1}$ with $k>l-1$. Then $C_{\mathcal{A}_n}$ is contained in the code generated by $A\otimes C_1$, which is the generator matrix for the matrix product code 
	$$[C_1\cdots C_1]A$$
	and then, by \cite[Theorem 2.2]{dmpc} we have the result. The other inequality follows in a similar way.
\end{IEEEproof}

\begin{ex}
	Consider  Example~\ref{hermit2} again  taking $\mathcal{A}_2=\{y_1x_2,y_1,x_2,x_1,1\}$. This code is contained in the decreasing code generated by $\mathcal{A}_2\cup\{x_2x_1\}$, then  $K_1=2>K_2=1$. We know that $m_3=3$ and $m_2=2$ and the minimum distances for these hermitian codes are  $3$ and $2$ respectively \cite{hermdist}, therefore
	$$\delta(C_{\mathcal{A}_2})\geq 6.$$
\end{ex}

Remember that isometry-dual condition for a sequence of codes $\{C_i\}_{i=1}^l$ means that exists $x\in\mathbb{F}_q^l$ such for each $i\in\{1,\ldots,l\}$, $C_i^\perp$ and $C_{l-i}$ are isometric according to $x$. Codes constructed from pointed curves $(\mathcal{X},Q)$ and $D\sim lQ$ satisfy this condition and we say that the curve satisfy the isometry-dual condition (\cite{diegorder}). We will see that polar codes constructed from these curves preserves a similar condition.

\begin{proposition}\label{dualpolar}
	Let $G$ be the kernel for a isometric-dual curve $(\mathcal{X},Q)$ of size $l\times l$. Let $C_{\mathcal{A}_n}$ be a decreasing code and define
	$$\mathcal{A}^\perp_n=\{ M^n_j\in\Delta(I_n)\ |\ j_i=l-1-k_i,\ 1\leq i\leq n,\ M^n_k\in\mathcal{A}_n\}^c$$
	Then $(C_{\mathcal{A}_n})^\perp$ is isometric to $C_{\mathcal{A}^\perp_n}$, and this code is also  decreasing.
\end{proposition}

\begin{IEEEproof}
	If we compare the sizes of the sets we just have to check that $C_{\mathcal{A}_n^\perp}\subset (C_{\mathcal{A}_n})^\perp$. It is also clear that $C_{\mathcal{A}_n^\perp}$ is also a decreasing code.
	Let $f(T)\in\mathcal{L}(\infty Q)$ be the element which establish the isometry between $C(\mathcal{X},D,m_iQ)^\perp$ and $C(\mathcal{X},D,m_{l-i}Q)$ for each $i\in\{1,\ldots,l\}$. Then we have that $\sum_{i=0}^{l-1} f(P_i)M_j(P_i)M_k(P_i)=0$ for each $j\in\{0,\ldots,l-1\}$ and for every $k\in\{0,\ldots,l-1-j\}$. Take $F=f(X_1)\cdots f(X_n)$ and $M^n_k\in\mathcal{A}_n$ and $M^n_{k'}\in C_{\mathcal{A}_n^\perp}$. Then we have
	$$\sum_{i=0}^{l^n-1}F(P^n_i)M^n_k(P^n_i)M^n_{k'}(P^n_i)=\prod_{j=1}^n \left(\sum_{i=0}^{l-1} f(P_i)M_{k_j}(P_i)M_{k'_j}(P_i)\right).$$
	We claim that there  exists $j\in\{1,\ldots,n\}$ such that $k'_j\leq l-1-k_j$. If this does not  happen we would  have
	$$k'_j>l-1-k_j,\ \forall j\in\{1,\ldots,n\}$$
	$$l-1-k'_j<k_j,\ \forall j\in\{1,\ldots,n\}$$
	 but $C_{\mathcal{A}_n}$ is decreasing, then for $\overline{k'}=\sum_{j=1}^n (l-1-k'_j)l^{j-1}$, $M^n_{\overline{k'}}\in\mathcal{A}_n\setminus\mathcal{A}_n^\perp$, which is a contradiction. Therefore it exists such $j$ and the sum over it is $0$ and we have the result.
\end{IEEEproof}

\begin{corollary}
	If in the  proof of Proposition~\ref{dualpolar} we have that the function $f$ evaluates to $ev(f)=(1,1,\ldots,1)$, then 
	$$C_{\mathcal{A}_n}^\perp=C_{\mathcal{A}_n^\perp}$$
	Codes with kernel $G_q$ satisfies this condition.
\end{corollary}

\begin{corollary}
	Let $C_{\mathcal{A}_n}$ be a decreasing code from a isometric-dual curve. Let $\mathcal{B}_n$ and $\mathcal{C}_n$ be decreasing sets such that
	$$\mathcal{C}_n\subset\mathcal{A}_n\subset\mathcal{B}_n$$
	then
	$$C_{\mathcal{B}^\perp_n}\subset C^\perp_{\mathcal{A}_n}\subset C_{\mathcal{C}^\perp_n}.$$
\end{corollary}

\begin{ex}
	We have already mentioned that each polar code can be seen as a shortened code obtained from a decreasing code, then we can complete its dual from the dual of the decreasing one. Let us take again
	$$\mathcal{A}_2=\{y_1x_2,x_2,y_1,x_1,1\}$$
	and 
	$$\mathcal{A}'_2=\mathcal{A}_2\cup\{x_1x_2\}.$$
	This is a decreasing set and we have
	\begin{align*}
	{\mathcal{A}'}^\perp_2=&\{x_1y_1x_2^3y_2, x_1^2x_2^3y_2, y_1x_2^3y_2, x_1x_2^3y_2,
	x_2^3y_2, x_1y_1x_2^2y_2, x_1^2x_2^2y_2, y_1x_2^2y_2,\\&
	x_1x_2^2y_2, x_2^2y_2, x_1^3y_1x_2^3, x_1^2y_1x_2^3, x_1^3x_2^3, x_1y_1x_2^3, x_1^2x_2^3, y_1x_2^3, x_1x_2^3, x_2^3,\\&
	x_1^3y_1x_2y_2, x_1^2y_1x_2y_2, x_1^3x_2y_2, x_1y_1x_2y_2, x_1^2x_2y_2, y_1x_2y_2, x_1x_2y_2, x_2y_2,\\&
	x_1^3y_1x_2^2, x_1^2y_1x_2^2, x_1^3x_2^2, x_1y_1x_2^2, x_1^2x_2^2, y_1x_2^2, x_1x_2^2, x_2^2, x_1^3y_1y_2, x_1^2y_1y_2,\\&
	x_1^3y_2, x_1y_1y_2, x_1^2y_2, y_1y_2, x_1y_2, y_2, x_1^3y_1x_2, x_1^2y_1x_2, x_1^3x_2, x_1y_1x_2, x_1^2x_2, y_1x_2,\\&
	x_1x_2, x_2, x_1^3y_1, x_1^2y_1, x_1^3, x_1y_1, x_1^2, y_1, x_1, 1\}
	\end{align*}
	In this case the isometry is given by $(1,\ldots,1)$ and if we add an orthogonal vector to the evaluations of $\mathcal{A}_2$ but not to the one of $x_1x_2$, we would have a generator set for $C^\perp_{\mathcal{A}_2}$. One of these vectors is the evaluation of $g=x_2x_1(x_1y_1+1)$, then ${\mathcal{A}'}_2^\perp\cup\{g\}$ generates $C^\perp_{\mathcal{A}_2}$.
\end{ex}

All the conditions asked for the pointed curves $(\mathcal{X},Q)$ are satisfied by weak Castle and Castle curves \cite{castillo1}. We say that a pointed curve $(\mathcal{X},Q)$ over $\mathbb{F}_q$ is weak Castle if $H(Q)$ is symmetric and there is a morphism $\phi:\overline{\mathbb{P}}\rightarrow\overline{\mathbb{P}}$ with $(\phi)_\infty=hQ$ and $\alpha_1,\ldots,\alpha_a\in\mathbb{F}_q$ such that
$$\left|\phi^{-1}(\alpha_i)\cap\mathcal{X}(\mathbb{F}_q)\right|=h.$$
$(\mathcal{X},Q)$ be a pointed Castle curve if it is weak Castle and $h$ is the multiplicity of $H(Q)$ and $r=q$.

\section{Modifying kernels from algebraic curves}

Remember that given a square-matrix $G$ of size $l\times l$ over $\mathbb{F}_q$ with rows $G_1,\ldots,G_l$, the exponent $E(G)$ of the matrix $G$ is defined as
$$E(G)=\frac{1}{l\ln l}\sum_{i=1}^l \ln D_i,$$
where $D_i$ is the called partial distance and it is the minimum of the Hamming distances $d(G_i,v)$, with $v\in\langle G_{i+1},\ldots, G_l\rangle$.

Suppose $G$ is non-singular over $\mathbb{F}_q$ of size $l\times l$ and $G'$ is as $G$. If ${G'}G^{-1}$ is a upper-triangular invertible matrix, then $E(G)=E(G')$; therefore, each matrix coming from a pointed curve has the same exponent. Looking for the best matrices over a given size, shortening codes is a good way to find them, for example this was the way to find the best matrix over $\mathbb{F}_2$ of size 16 (see \cite{korada}).

Next theorem was proved by Anderson and Matthews in \cite{Anderson}. It says that shortening kernels from algebraic curves does not change the final structure of the code.

\begin{theorem}\label{qpag}
	Let $G$ be a kernel from a pointed curve $(\mathcal{X},Q)$ with $D=\sum_{i=1}^l P_i$. Taking the $j$-th column, we can shorten $G$ to obtain the matrix $G'$. Then we have that $G'$ is the kernel arising from the codes $\{C(D-P_j,mQ-P_j)\}_{m\in H^\ast(Q)}$.
\end{theorem}
We can repeat this process to obtain polar codes from kernel associated to divisors of the form $mQ-\sum P$. However, if we take points coming from zero divisor of elements in $\mathcal{L}(\infty Q)$ we will have a matrix with the same structure.

\begin{proposition}
	Let $(\mathcal{X},Q)$ a pointed curve and $z\in\mathbb{F}_q(\mathcal{X})$ with $(z)=D-lQ$, $D=\sum_{i=1}^l P_i$. Let's suppose there is $z'\in\mathbb{F}_q(\mathcal{X})$ such that $(z')=\sum_{i=1}^s P_{k_i}-sQ$ with $k_i\neq k_j$ if $i\neq j$; define $D'=(z')_0$.
	
	Let $\varphi:\mathbb{F}_q^l\rightarrow\mathbb{F}_q^s$ be the mapping such $\varphi(c)$ is the same word $c$ but erasing the entries indexed by $\{i\in\{1,\ldots,n\}\ |\ P_i\notin\mathrm{supp}\ z'\}$. Let $\psi:R/\langle I,f_z\rangle\rightarrow R/\langle I,f_{z'}\rangle$ the natural mapping between both rings. Then
	\[\begin{tikzcd}
	\arrow[d, "\psi"]R/\langle I,f_z\rangle \arrow[r, "ev_D"]  &\mathbb{F}_q^n  \arrow[d, "\varphi"]\\
	R/\langle I,f_{z'}\rangle 
	\arrow[r,"ev_{D'}"]& \mathbb{F}_q^{s}
	\end{tikzcd}\]
	is commutative. The kernel $G$ constructed from $D$ and $Q$ has as submatrix $G'$, the kernel from $D'$ and $Q$.
\end{proposition}

\begin{IEEEproof}
	Take $f,f'\in R/\langle I,f_z\rangle$ such that
	$$\varphi(ev_D(f))=\varphi(ev_D(f')).$$
	This occurs if and only if
	$$ev_D(f)_j=ev_D(f')_j\ \ \forall j\ \in P_j\in\mathrm{supp}\ z'.$$
	This is $ev_{D'}(f)=ev_{D'}(f')$, then we have $f-f'\in\langle I,f_{z'}\rangle$, implying $\psi(f)=\psi(f')$ as we wanted it. 
\end{IEEEproof}

\begin{corollary}
	The matrix $G'$ of the proposition above is isometric to the one obtained after shortening $G$ with the process described in Theorem \ref{qpag}.
\end{corollary}

\medskip

\begin{corollary}
	
	A Castle-like curve with $D=(z)_0=\left(\prod_{i=1}^a (\phi-\alpha_i)\right)_0$ produces a sequence of $a$ kernels (each one submatrix of the next) coming from the divisors of $\prod_{i=1}^j (\phi-\alpha_i)$, $j\in\{1,\ldots,a\}$. 
\end{corollary}

\medskip

\begin{ex}
	Take again the hermitian curve over $\mathbb{F}_4$ where $\alpha$ is a primitive element, $x^3=y^2+y$. This is a Castle curve with kernel
	$$\begin{array}{r|cccccccc}
	
	& 00 & 01 & 1\alpha & 1\alpha^2& \alpha\alpha & \alpha\alpha^2 & \alpha^2\alpha & \alpha^2\alpha^2\\\hline
	
	x^3y&0&0&\alpha&\alpha^2&\alpha&\alpha^2&\alpha&\alpha^2\\
	
	x^2y & 0& 0&\alpha&\alpha^2&1&\alpha&\alpha^2&1\\
	
	x^3& 0&0&1&1&1&1&1&1\\
	
	xy & 0&0&\alpha&\alpha^2&\alpha^2&1&1&\alpha\\
	
	x^2& 0&0&1&1&\alpha^2&\alpha^2&\alpha&\alpha\\
	
	y&0&1&\alpha&\alpha^2&\alpha&\alpha^2&\alpha&\alpha^2\\
	
	x&0&0&1&1&\alpha&\alpha&\alpha^2&\alpha^2\\
	
	1&1&1&1&1&1&1&1&1
	\end{array}$$
	If we shorten this kernel taking the points with $x=0$ like in Theorem \ref{qpag}, starting with $00$.
	$$\begin{array}{r|cccccc}
	
	& 1\alpha & 1\alpha^2& \alpha\alpha & \alpha\alpha^2 & \alpha^2\alpha & \alpha^2\alpha^2\\\hline
	
	x^3y&\alpha&\alpha^2&\alpha&\alpha^2&\alpha&\alpha^2\\
	
	x^2y &\alpha&\alpha^2&1&\alpha&\alpha^2&1\\
	
	x^3&1&1&1&1&1&1\\
	
	xy &\alpha&\alpha^2&\alpha^2&1&1&\alpha\\
	
	x^2&1&1&\alpha^2&\alpha^2&\alpha&\alpha\\

	x&1&1&\alpha&\alpha&\alpha^2&\alpha^2
	\end{array}.$$
	This matrix comes from the codes with divisor $(x^3-1)_0$ and $P_\infty-P_{00}-P_{01}$.
	From the original kernel, if we remove the columns of that points and the rows products of $x^3$.
	
	$$\begin{array}{r|cccccccc}
	
	& 1\alpha & 1\alpha^2& \alpha\alpha & \alpha\alpha^2 & \alpha^2\alpha & \alpha^2\alpha^2\\\hline

	x^2y &\alpha&\alpha^2&1&\alpha&\alpha^2&1\\
	
	xy &\alpha&\alpha^2&\alpha^2&1&1&\alpha\\
	
	x^2&1&1&\alpha^2&\alpha^2&\alpha&\alpha\\
	
	y&\alpha&\alpha^2&\alpha&\alpha^2&\alpha&\alpha^2\\
	
	x&1&1&\alpha&\alpha&\alpha^2&\alpha^2\\
	
	1&1&1&1&1&1&1
	\end{array}$$
	
	This matrix comes from the divisor $(x^3-1)_0$ and $P_\infty$. The isometry between both matrices is clear and this second matrix has the same structure as the original one, so we can apply the analysis of information set, minimum distance and its dual like before.
	
	As an example of the last corollary we can give the next matrix sequence from the hermitian curve
	$$\begin{array}{r|cc}
	&00&01\\\hline
	y&0&1\\
	1&1&1\end{array}\ \ \ \begin{array}{r|cccc}
	&00&01&1\alpha&1\alpha^2\\\hline
	xy&0&0&\alpha&\alpha^2\\
	y&0&1&\alpha&\alpha^2\\
	x&0&0&1&1\\
	1&1&1&1&1\end{array}
	\ \ \begin{array}{r|cccccc}
	&00&01&1\alpha&1\alpha^2&\alpha\alpha&\alpha\alpha^2\\\hline
	x^2y&0&0&\alpha&\alpha^2&1&\alpha\\
	xy&0&0&\alpha&\alpha^2&\alpha^2&1\\
	x^2&0&0&1&\alpha^2&\alpha^2&\alpha\\
	y&0&1&\alpha&\alpha^2&\alpha&\alpha^2\\
	x&0&0&1&1&\alpha&\alpha\\
	1&1&1&1&1&1&1\end{array}
	$$
	
	$$
	\begin{array}{r|cccccccc}
	
	& 00 & 01 & 1\alpha & 1\alpha^2& \alpha\alpha & \alpha\alpha^2 & \alpha^2\alpha & \alpha^2\alpha^2\\\hline
	
	x^3y&0&0&\alpha&\alpha^2&\alpha&\alpha^2&\alpha&\alpha^2\\
	
	x^2y & 0& 0&\alpha&\alpha^2&1&\alpha&\alpha^2&1\\
	
	x^3& 0&0&1&1&1&1&1&1\\
	
	xy & 0&0&\alpha&\alpha^2&\alpha^2&1&1&\alpha\\
	
	x^2& 0&0&1&1&\alpha^2&\alpha^2&\alpha&\alpha\\
	
	y&0&1&\alpha&\alpha^2&\alpha&\alpha^2&\alpha&\alpha^2\\
	
	x&0&0&1&1&\alpha&\alpha&\alpha^2&\alpha^2\\
	
	1&1&1&1&1&1&1&1&1
	\end{array}.$$
	
	Their exponents are, respectively
	$$\frac{1}{2},\ \frac{1}{2},\ \frac{\ln(6\cdot 4\cdot 3\cdot 2\cdot 2\cdot 1)}{6\ln(6)}\approx 0.5268,\ \frac{\ln(8\cdot 6\cdot 5\cdot 4\cdot 3\cdot 2\cdot 2\cdot 1)}{8\ln (8)}\approx 0.5622$$ 
\end{ex}

Now we will check another resource to search matrices with good exponents arising from AG codes. We will need the next result.

\begin{proposition}
	Let $G$ and $G'$ be two matrices over $\mathbb{F}_q$ of size $l\times l$ and $l'\times l'$ respectively, non-singular and with partial distances $\{D_i(G)\}_{i=1}^l$ and $\{D_i(G')\}_{i=1}^{l'}$. Then for the matrix $G'\otimes G$ we have
	$$D_k(G'\otimes G)=D_{i'}(G')\cdot D_i(G)$$
	where $k=(i'-1)l+i$.
\end{proposition}

\begin{IEEEproof}
	For the first $l$ rows is clear since they are just copies of the original $G$. Let us suppose the result for the first $l'l-(kl+l)+1$ rows ($0\leq k\leq l'-1$) and let's prove it for the rows $l'l-(k+1)l-j$, $0\leq j\leq l-1$.
	
	If we  begin with  $h=l'l-(k+1)l$ we observe that the partial distance $D_h(G'\otimes G)$ is the same as the one of
	\begin{equation}
	\begin{bmatrix} \{G'_i\}_{i=1}^k\otimes G\\
	\{G'_i\}_{i=k+1}^{l'}\otimes I_l\end{bmatrix}\tag{$\ast$}
	\end{equation}	
	where $I_l$ is the identity matrix of size $l$ and $G'_i$ is the $i$-th row of $G'$. Since the matrix $G$ is non-singular and the $h$-th partial distance this is the distance $d(G'\otimes G_h,\langle G'\otimes G_{h+1},\ldots,G'\otimes G_{l'l}\rangle)$, then the last vector space is generated by the tensor product of the last $l'-k$ rows of $G'$ with $I_l$. Also we know that
	$$G'_{l'-k}\otimes G_l=\sum_{j=1}^l G'_{l'-k}\otimes (G_{l,j}e_j),$$
	 where $e_j$ is the $j$-th vector of the canonical basis for $\mathbb{F}_q^l$. Notice that if $u$ and $v$ are two vectors with disjoint supports, then the Hamming weight $w(v+u)=w(v)+w(u)$; also, $w(v\otimes u)=w(v)\cdot w(u)$. Then if we take some elements $\alpha_i\in\mathbb{F}_q$, $l'l-(k+1)l+1\leq i\leq l'l$ we have
	\begin{align*}
	&w\left(G'_{l'-k}\otimes G_l+\sum_{i=l'-k+1,j=1}^{l',l},\alpha_{(i-1)l+j}G'_i\otimes e_j\right)\\
	=&w\left(\sum_{j=1}^l G'_{l'-k}\otimes G_{l,j}e_j+\sum_{i=l'-k+1,j=1}^{l',l}\alpha_{(i-1)l+j}G'_i\otimes e_j\right)\\
	=&w\left(\sum_{j\in\mathrm{supp}\ G_l}\left(G'_{l'-k}+\sum_{i=l'-k+1}^{l'}\frac{\alpha_{(i-1)l+j}}{G_{l,j}}G'_i\right)\otimes e_j+\sum_{j\notin\mathrm{supp}\ G_l}\left(\sum_{i=l'-k+1}^{l'} \alpha_{(i-1)l+j}G'_i\right)\otimes e_j\right)\\
	\stackrel{\circ}{=}&\sum_{j\in\mathrm{supp}\ G_l}w\left(\left(G'_{l'-k}+\sum_{i=l'-k+1}^{l'}\frac{\alpha_{(i-1)l+j}}{G_{l,j}}G'_i\right)\otimes e_j\right)+\sum_{j\notin\mathrm{supp}\ G_l}w\left(\left(\sum_{i=l'-k+1}^{l'} \alpha_{(i-1)l+j}G'_i\right)\otimes e_j\right)\\
	\geq&\sum_{j\in\mathrm{supp}\ G_l}w\left(\left(G'_{l'-k}+\sum_{i=l'-k+1}^{l'}\frac{\alpha_{(i-1)l+j}}{G_{l,j}}G'_i\right)\otimes e_j\right)\\
	=&\sum_{j\in\mathrm{supp}\ G_l}w\left(G'_{l'-k}+\sum_{i=l'-k+1}^{l'}\frac{\alpha_{(i-1)l+j}}{G_{l,j}}G'_i\right)w(e_j)\\
	\geq &\sum_{j\in\mathrm{supp}\ G_l} D_{l'-k}(G')\\
	=&D_l(G)\cdot D_{l'-k}(G'),
	\end{align*}
	where we have that $\circ$ sin $v\otimes e_j$ has disjoints support for different $j$.
	The results follows if we change $G_l$ for any vector of weight $D_j(G)$. 
\end{IEEEproof}

Next corollary is a general version of the one in \cite{Kron}.

\begin{corollary}\label{expkron}
	Let $G_1$ and $G_2$ be two non-singular matrices over $\mathbb{F}_q$ of sizes $l_1$ and $l_2$ respectively. Then
	$$E(G_1\otimes G_2)=\frac{E(G_1)}{\log_{l_1}(l_1l_2)}+\frac{E(G_2)}{\log_{l_2}(l_1l_2)}.$$
\end{corollary}

\begin{IEEEproof}
	We know that $G_1\otimes G_2$ has size $l_1l_2$. For each $k\in\{1,\ldots,l_1l_2\}$ we can rewrite $k$ as $k=(j-1)l_2+s$ where $1\leq j\leq l_1$ y $1\leq s\leq l_2$, therefore
	\begin{align*}
	E(G_1\otimes G_2)&=\frac{1}{l_1l_2\ln(l_1l_2)}\sum_{k=1}^{l_1 l_2}\ln(D_k(G_1\otimes G_2))\\
	&\stackrel{(a)}{=}\frac{1}{l_1l_2\ln(l_1l_2)}\sum_{j=1}^{l_1}\sum_{s=1}^{l_2}\ln(D_j(G_1)D_s(G_2))\\
	&=\frac{1}{l_1\ln(l_1l_2)}\sum_{j=1}^{l_1} \ln(D_j(G_1))+\frac{1}{l_2\ln(l_1l_2)}\sum_{s=1}^{l_2}\ln(D_s(G_2))\\
	&=\frac{E(G_1)\ln l_1}{\ln l_1l_2}+\frac{E(G_2)\ln l_2}{\ln l_1l_2}
	\end{align*}
	where equality $(a)$ follows from the previous proposition.
\end{IEEEproof}

We can extend the analysis done for the kernel defined by one curve to the product of two kernels arising from two curves defined over the same field.
Let  $\left(\mathcal{X},(z)_0=\sum_{i=0}^{l_1-1} P_i,Q\right)$ and $\left(\mathcal{Y},(z')_0=\sum_{j=0}^{l_2-1}P'_i,Q'\right)$ be two pointed curves over $\mathbb{F}_q$ and $S[X]=\mathbb{F}_q[x_1,\ldots,x_s]$ and $S[Y]=\mathbb{F}_q[y_1,\ldots,y_{s'}]$ the polynomial rings where there exist   $I\subset S[X]$ and $I'\subset S[Y]$ such that $S[X]/\langle I,f_z\rangle$ and $S[Y]/\langle I',f_{z'}\rangle$ are isomorphic to the codes associated to the respective curves.We will denote as  $G_X$ and $G_Y$ their respective kernels. 

 We denote by $S[X,Y]=\mathbb{F}_q[x_1,\ldots,x_s,y_1,\ldots,y_{s'}]$ and by $I_{XY}=\langle I, I',f_z,f_{z'}\rangle$. We will endow $R[X,Y]=S[X,Y]/\langle I,I'\rangle$ with the weight generated by the inherit vectors $$w_1=(w(x_1),\ldots,w(x_s),0,\ldots,0)\hbox{ and } w_2=(0,\ldots,0,w(y_1),\ldots,w(y_{s'})).$$

\begin{proposition}
	If $\Delta(I)=\{M_0,\ldots, M_{l_1-1}\}$ and $\Delta(I')=\{M'_0,\ldots,M'_{l_2-1}\}$, then 
	$$\Delta(I_{XY})=\{M_0M'_0,M_0M'_1,\ldots,M_{l_1-1}M_{l_2-1}\}$$
 and the rows of the matrix $G_X\otimes G_Y$ are evaluations of the elements in  $\Delta(I_{XY})$
	$$M\cdot M'(Q_i)=M(P_j)M'(P'_k)\ \ \ i=jl_2+k$$
	in decreasing order w.r.t. the induced ordering.
\end{proposition}

\begin{IEEEproof}
	The equality for  $\Delta(I_{XY})$ is clear and the equality on the rows follows from the definition of the Kronecker product since
		\begin{align*}
	(G_X\otimes G_Y)_{i,j}&=(G_X)_{\lfloor i/l_2\rfloor,\lfloor j/l_2\rfloor}(G_Y)_{i\ \mathrm{mod}\ l_2,j\ \mathrm{mod}\ l_2}\\
	&=M_{l^n-\lfloor i/l_2\rfloor}(P_{\lfloor j/l_2\rfloor})M'_{l^n-i\ \mathrm{mod}\ l_2}(P'_{j\ \mathrm{mod}\ l_2}).
	\end{align*}
\end{IEEEproof}

Thus we have a set of monomials  $\tilde{M}_i=M_{\lfloor i/l_2\rfloor}M'_{i\mathrm{mod}\ l_2}$ and $Q_i$ as before. Let $l=l_1l_2$ and consider the polar code constructed from the kernel  $G_{XY}$. Now we will work on the polynomial ring $R[X_1,Y_1,X_2,Y_2,\ldots,X_n,Y_n]$.

We will define an ordering on  $\mathbb{Z}$ as follows $i\triangleleft j$ if and only if  when $i=hl_2+k$ and $j=h'l_2+k'$ we have $h<h'$ y $j<j'$. With this new ordering our previous definitions are translated easily.

\begin{itemize}
	\item A code $C_{\mathcal{A}_n}$ with kernel $G_{XY}$ is called weakly decreasing if for  $\tilde{M}^n_k\in\mathcal{A}_n$, $\tilde{M}^n_{k'}|\tilde{M}^n_k$ it follows that $\tilde{M}^n_{k'}\in\mathcal{A}_n$. As a corollary a polar code over a  SOF channel is  weakly decreasing.
	
	\item A code is decreasing if it is weakly decreasing  and also $\tilde{M}_{i_1}(X_{j_1}Y_{j_1})\cdots\tilde{M}_{i_k}(X_{j_k}Y_{j_k})\in\mathcal{A}_n$ implies
	$$\tilde{M}_{i_1}(X_{j'_1}Y_{j'_1})\cdots\tilde{M}_{i_k}(X_{j'_k}Y_{j'_k})\in\mathcal{A}_n$$
	for $j'_v\leq j_v$, $v\in\{1,\ldots,k\}$.  As before this last property will be call degrading property.
\end{itemize}

\begin{ex} From Corollary~\ref{expkron} we can see that a matrix $G^{\otimes n}$ has the same exponent as the original matrix $G$.
Let us consider the field $\mathbb{F}_4$ and the field of rational functions with variable $t$ and the hermitian curve over $\mathbb{F}_4[x,y]$. We compute the kernel using $Q=P_\infty$ (the common pole of  $x$ and $y$) and we construct the matrices  $G_H$ and $G_R$.  The monomial basis we have to consider are
	$$L_H=\{x^3y,x^2y,x^3,xy,x^2,x,y,1\},$$
	$$L_R=\{t^3,t^2,t,1\}.$$
	Therefore  $G_H\otimes G_R$ is the evaluation of the monomials 
	\begin{align*}
	L_{H\otimes R}=&\{x^3yt^3,x^3yt^2,x^3yt,x^3y,x^2yt^3,x^2yt^2,x^2yt,x^2y,\\
	&x^3t^3,x^3t^2,x^3t,x^3,xyt^3,xyt^2,xyt,xy,\\ &x^2t^3,x^2t^2,x^2t,x^2,yt^3,yt^2,yt,y,\\
	&xt^3,xt^2,xt,x,t^3,t^2,t,1\}.
	\end{align*}
	The rational curve has exponent $\frac{\ln 4!}{4\ln 4}$ and the hermitian one  has exponent $\frac{\ln(8\cdot 2\cdot 6!)}{8\ln 8}$ thus the kernel  $G_H\otimes G_R$ has exponent
	$$E(G_H\otimes G_R)\approx 0.5665\,.$$
	If we construct a polar code from this kernel over a SOF channel and $n=1$ we have that if $xyt^2,x^2t^2\in\mathcal{A}_n$ then
	$$xyt,xy,x^2t,x^2,yt^2,yt,y,xt^2,xt,x,t^2,t,1\in\mathcal{A}_n.$$
	If those are the only elements in the code then we have a decreasing code with minimum distance $6$ (since the code associated to $xy$ has minimum distance 3 and the one associated to  $t^2$ has minimum distance 2).
\end{ex}
\section{Conclusion}
In this paper we have stablished a construction of polar codes  from pointed algebraic curves for  a
discrete memoryless channel which is symmetric w.r.t the field operations. This results extend some results in \cite{Bardet} for  a binary symmetric channel.  Note that both the families of weak Castle and Castle curves provide good candidates for designing the proposed polar codes since they satisfy  the conditions needed in the construction. Even if the nature of the results is mainly theoretical, we believe that it can contribute to a deeper understanding to polar codes over non-binary alphabeths.


\begin{thebibliography}{99}

\bibitem{Anderson} Anderson, S. E., \& Matthews, G. L. (2014). Exponents of polar codes using algebraic geometric code kernels. Designs, codes and cryptography, 73(2), 699-717.

\bibitem{Arikan} Ar\i kan, E. (2009). Channel polarization: A method for constructing capacity-achieving codes for symmetric binary-input memoryless channels. IEEE Transactions on Information Theory, 55(7), 3051-3073.

\bibitem{Bardet} Bardet, M., Dragoi, V., Otmani, A., \& Tillich, J. P. (2016). Algebraic properties of polar codes from a new polynomial formalism. In International Symposium on Information Theory ISIT 2016 (pp. 230-234).


\bibitem{dpi}  Cover, T. M., \& Thomas, J. A. (2012). Elements of information theory. John Wiley \& Sons.


\bibitem{qSC3} Bleichenbacher, D., Kiayias, A., \& Yung, M. (2003, June). Decoding of interleaved Reed Solomon codes over noisy data. In International Colloquium on Automata, Languages, and Programming (pp. 97-108). Springer, Berlin, Heidelberg.

\bibitem{qSC4} Dodunekova, R., Dodunekov, S. M., \& Nikolova, E. (2008). A survey on proper codes. Discrete Applied Mathematics, 156(9), 1499-1509.

\bibitem{diegorder} Geil, O., Munuera, C., Ruano, D., \& Torres, F. (2010). On the order bounds for one-point AG codes. arXiv preprint arXiv:1002.4759.

\bibitem{ruudorder} Geil, O., \& Pellikaan, R. (2002). On the structure of order domains. Finite Fields and Their Applications, 8(3), 369-396.

\bibitem{dmpc} Hernando, F., Lally, K., \& Ruano, D. (2009). Construction and decoding of matrix-product codes from nested codes. Applicable Algebra in Engineering, Communication and Computing, 20(5-6), 497.

\bibitem{korada} Korada, S. B., \c Sa\c so\u glu, E., \& Urbanke, R. (2010). Polar codes: Characterization of exponent, bounds, and constructions. IEEE Transactions on Information Theory, 56(12), 6253-6264. 

	\bibitem{qSC2} Lechner, G., \& Weidmann, C. (2008, September). Optimization of binary LDPC codes for the q-ary symmetric channel with moderate q. In Turbo Codes and Related Topics, 2008 5th International Symposium on (pp. 221-224). IEEE.
	
\bibitem{Kron} Lee, M. K., \& Yang, K. (2014). The exponent of a polarizing matrix constructed from the Kronecker product. Designs, codes and cryptography, 70(3), 313-322.

\bibitem{AGC} Mart\'{i}nez-Moro, Edgar and Munuera, Carlos and Ruano, Diego (Eds.) (2008). Advances in algebraic geometry codes. Series on Coding Theory and Cryptology Vol. 5. World Scientific Publishing Co. Pte. Ltd.

\bibitem{morinl} Mori, R., \& Tanaka, T. (2010). Channel polarization on q-ary discrete memoryless channels by arbitrary kernels. In Information Theory Proceedings (ISIT), 2010 IEEE International Symposium on (pp. 894-898). IEEE.

\bibitem{mori2} Mori, R.,  \& Tanaka, T. (2010). Non-binary polar codes using Reed-Solomon codes and algebraic geometry codes. In Information Theory Workshop (ITW), 2010 IEEE (pp. 1-5). IEEE.

\bibitem{moriq} Mori, R., \& Tanaka, T. (2014). Source and channel polarization over finite fields and Reed–Solomon matrices. IEEE Transactions on Information Theory, 60(5), 2720-2736.

\bibitem{castillo1} Munuera, C., Sep\'ulveda, A., \& Torres, F. (2008). Algebraic Geometry codes from Castle curves. In Coding Theory and Applications (pp. 117-127). Springer, Berlin, Heidelberg.

\bibitem{Sasoglu} \c Sa\c so\u glu, E., Telatar, E., \& Ar\i kan, E. (2009, October). Polarization for arbitrary discrete memoryless channels. In Information Theory Workshop, 2009. ITW 2009. IEEE (pp. 144-148)

\bibitem{qSC1} Shokrollahi, A. (2004, October). Capacity-approaching codes on the q-ary symmetric channel for large q. In Information Theory Workshop, 2004. IEEE (pp. 204-208). IEEE.


\bibitem{hermdist} Yang, K., \& Kumar, P. V. (1992). On the true minimum distance of Hermitian codes. In Coding theory and algebraic geometry (pp. 99-107). Springer, Berlin, Heidelberg.

\end{thebibliography}
\end{document}